\documentclass[12pt]{iopart}
\usepackage{graphicx}

\begin{document}

\title[]{Superfluid current disruption in a chain of weakly
coupled Bose-Einstein Condensates}

\author{F.~S.~Cataliotti\dag\ \footnote[3]{To whom correspondence
should be addressed (fsc@lens.unifi.it)}, L.~Fallani, F.~Ferlaino,
C.~Fort, P.~Maddaloni\ddag, M. Inguscio}

\address{LENS, Dipartimento di Fisica, Universit\`a di Firenze and INFM, via Nello Carrara 1,
I-50019 Sesto Fiorentino (Firenze), Italy}

\address{\dag\ also INFM and Dipartimento
di Fisica, Universit\`a di Catania, via S. Sofia~64, I-95124
Catania, Italy}

\address{\ddag\ also Istituto Nazionale di Ottica Applicata
(INOA), via Campi Flegrei 34, I-80078 Pozzuoli (Napoli), Italy }

\begin{abstract}
We report the experimental observation of the disruption of the
superfluid atomic current flowing through an array of weakly
linked Bose-Einstein condensates. The condensates are trapped in
an optical lattice superimposed on a harmonic magnetic potential.
The dynamical response of the system to a change of the magnetic
potential minimum along the optical lattice axis goes from a
coherent oscillation (superfluid regime) to a localization of the
condensates in the harmonic trap (``classical" {\em insulator}
regime). The localization occurs when the initial displacement is
larger than a critical value or, equivalently, when the velocity
of the wavepacket's center of mass is larger than a critical
velocity dependent on the tunnelling rate between adjacent sites.
\end{abstract}

\maketitle

Atomic Bose-Einstein condensates have been either loaded or
produced in periodic potentials opening up the possibility to
investigate new phenomena tuning the degree of coherence in the
system. Experiments have explored regimes ranging from the
coherent matter wave emission from a condensate loaded on a
vertical standing wave \cite{anderson98}, to the observation of
number squeezed states \cite{orzel01}, to the demonstration of a
one-dimensional Josephson junctions array with a linear chain of
condensates produced in an optical lattice \cite{cataliotti01}, to
the recent observation of a quantum phase transition in a
condensate loaded in a $3D$ optical lattice \cite{greiner02}. The
dynamical behavior of coherent matter waves in periodic potentials
has also been the subject of extensive theoretical work. Different
phenomena have been predicted including the formation of bright
solitons and the appearance of various forms of instabilities
where the usual long range phase coherence properties are lost
\cite{trombettoni01,bronski01,carusotto02,smerzi02,wu01,cardenas02,adhikari}.

In this paper we report on the experimental observation of the
disruption of superfluidity in a linear array of $^{87}$Rb
condensates trapped in the combined potential of an optical
standing wave superimposed on a harmonic magnetic trap.

We observe a transition from a regime in which the wavepacket
coherently oscillates in the array to another one in which the
condensates stop in the optical potential sites and lose their
relative phase coherence. We also observe the appearance of an
intermediate regime where no oscillations are visible while the
relative phase coherence is not completely lost. In a previous
experiment we have investigated the regime of small periodic
potential height, showing that in this regime and for large atom
numbers, the mechanism for superfluidity breakdown was in
agreement with predictions based on the onset of the Landau
instability \cite{sven}. In the present experiment we are
exploring the dynamics of the system in the case of a large
periodic potential height, where a modulational instability is
expected to play a fundamental role \cite{smerzi02,wu01}.
Furthermore, one can study the overall coherence of the system
observing the phase evolution in the interferogram of the expanded
array of condensates.

The experimental setup has been already described in
\cite{cataliotti01}: a dilute vapor of $^{87}Rb$ atoms confined in
a Ioffe-type magnetic trap in the ($F=1$, $m_F=-1$) state is Bose
condensed by the rf-evaporative cooling. Slightly above the
condensation threshold, we superimpose to the harmonic magnetic
trap a $1D$ optical standing wave. We then continue the
evaporation process through the phase-transition temperature. The
magnetic trap is cigar-shaped, with frequencies
$\omega_z=2\pi\times 9$~Hz and $\omega_r=2 \pi\times 90$~Hz along
the axial and radial directions, respectively. The optical
standing wave is superimposed to the magnetic potential, along the
$z$-axis, and is created with a retroreflected collimated laser
beam, detuned $\sim 3$~nm to the blue of the rubidium D1
transition at $\lambda=795$~nm. The resulting periodic potential
is $V_L=V_0 \cos^2{(2 \pi z / \lambda)}$: the valleys of the
potential are separated by $\lambda/2$ and the interwell energy
barrier $V_0=sE_R$ can be controlled by varying the intensity of
the laser beam. $E_R=h^2 /2 m \lambda^2$ is the kinetic energy of
an atom of mass $m$ recoiling after absorbing one lattice photon.
Typically, in our experiments, $s$ ranges from 3 to 14~$E_R$, for
such values, the condensate chemical potential $\mu$, that ranges
from 1.8 to 3.7~$E_R$, is smaller than $V_0$ and the system
realizes an array of weakly coupled condensates driven by an
external harmonic field.

We drive the system out of equilibrium by a sudden displacement
($t_{dis}\ll 2\pi/ \omega _{z}$) of the magnetic potential along
the lattice axis. The ensuing dynamics is revealed by turning off,
after different evolution times, both the magnetic and optical
traps and imaging the atomic density distribution after an
expansion time $t_{exp}=27.8$~ms. During the expansion the
different condensates, originally located at the nodes of the
laser standing wave, overlap and interfere allowing to monitor
their relative phases \cite{pedri01}.

For small displacements we observe a superfluid regime that has
been extensively studied both at zero temperature
\cite{cataliotti01} and in the presence of a significant thermal
fraction \cite{ferlaino02}. The condensates released from the
traps show an interference pattern, consisting of three peaks,
separated by a distance $d=2 h t_{exp}/m \lambda$, oscillating in
phase, thus revealing the long-range coherence of the condensates
across the entire optical lattice.

\begin{figure}[h]
\begin{center}
\includegraphics[width=10cm]{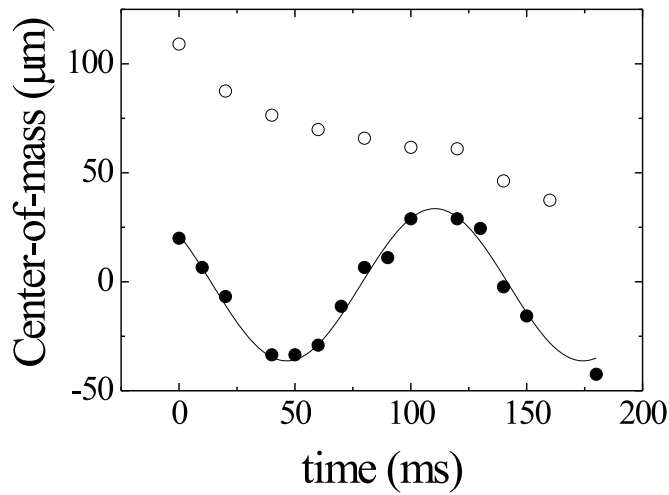}
\caption{Position of the center-of-mass of the atomic cloud in the
trap for an optical barrier height of 5~E$_R$. Filled circles:
displacement of 30~$\mu$m; open circles: displacement of
120~$\mu$m. The continuous line represents a fit to the ``small
displacement'' data with a sine function.} \label{fig1}
\end{center}
\end{figure}

When we increase the displacement of the magnetic trap center for
a fixed height of the optical barriers we observe a different
behaviour: the center of mass of the atomic sample no longer
oscillates but slowly moves towards the center of the magnetic
potential where it subsequently stops. This is shown in
Fig.~\ref{fig1} where we report the evolution of the
center-of-mass position of the atomic cloud inside the combined
trap for a barrier height of 5~E$_R$ and displacements of
30~$\mu$m and 120~$\mu$m. We reconstruct the center of mass motion
in the trap from the images of the expanded cloud assuming
ballistic expansion. We take data points at intervals of time
small enough so as to assume the condensate moving in the combined
trap with constant velocity during each time interval.

For the larger displacement we also observe a vanishing of the
visibility of the interference peaks demonstrating the loss of
long-range coherence across the array.

\begin{figure}[h]
\begin{center}
\includegraphics[width=10cm]{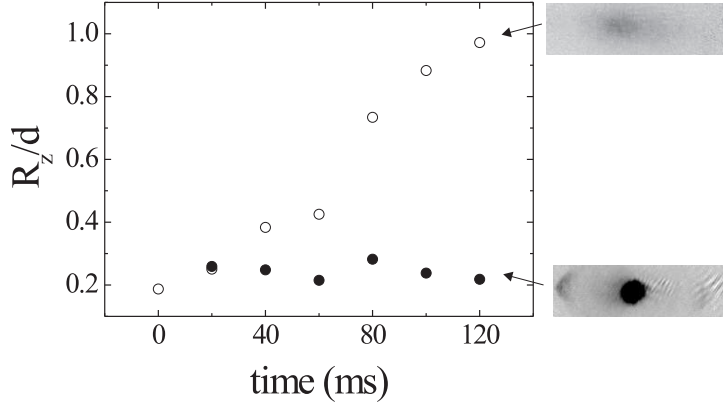}
\caption{Axial width ($R_z$) of the interferogram central peak
normalized to the peaks separation $d$ as a function of time spent
in the combined trap after the magnetic trap displacement for an
optical barrier height of 5~E$_R$.  Filled circles: displacement
of 30~$\mu$m; open circles: displacement of 120~$\mu$m. In the
right part of the figure we show the observed interferogram at
t=120~ms for the two displacements.} \label{fig2}
\end{center}
\end{figure}

In Fig.~\ref{fig2} we report the evolution of the axial width
$R_z$ of the interferogram central peak normalized to the peaks
separation $d$ for the same conditions as in Fig.~\ref{fig1}.
While in the small amplitude coherent oscillation the axial width
remains constant, in the larger displacement case it spreads out
reaching the peak separation in $\sim$120~ms when the
interferogram is completely washed out. We remark that the
vanishing of the interferogram is a dynamical effect and is not
due to scattering of lattice photons. Indeed in the absence of
displacement or even for small displacements we do not observe a
deterioration of the interference contrast on the time scales of
the experiment as shown in the right part of Fig.~\ref{fig2}.

\begin{figure}[h]
\begin{center}
\includegraphics[width=10cm]{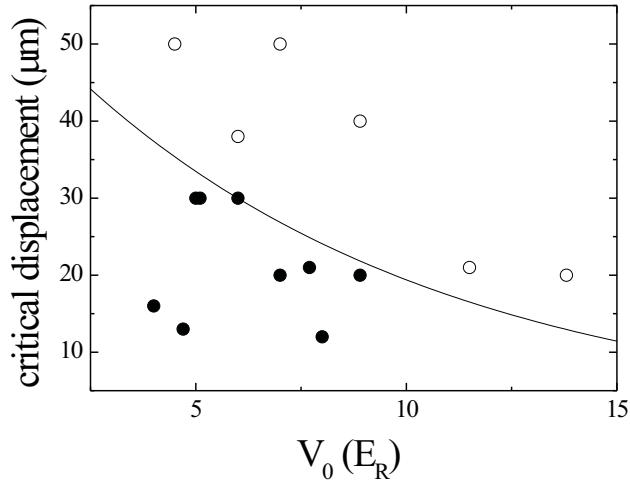}
\caption{Comparison of experimental results with
Eq.~(\ref{cr-displ}). Filled circles represent coherent
oscillations (with coherent interferogram featuring the three
peaks), empty circles denote pinned motions; continuous line
results of Eq.~(\ref{cr-displ}). } \label{fig3}
\end{center}
\end{figure}

We have studied the crossover between this two dynamical regimes
repeating the experiment varying the displacement for different
optical lattice heights $V_0$. In Fig.~\ref{fig3} we summarize the
experimental results: the closed circles correspond to values of
displacement and lattice depth for which we observe a coherent
oscillation. The open circles identify conditions where we have
the destruction of the oscillation and a slow, overdamped, motion
toward the center of the magnetic trap. In the latter case, only
for very large displacements we observe a complete vanishing of
the interferogram.

We compare our results with the prediction for the onset of the
dynamical instability recently proposed in \cite{smerzi02} using
the one dimensional Discrete Non-Linear Schr\"{o}dinger equation
(DNLS) model \cite{trombettoni01}. This treatment gives an
analytical expression for the ``critical'' velocity $v_{cr}$ to
enter the dynamical instability regime. The $1D$ system is
predicted to undergo a sudden transition from a regime with
coherent oscillations ({\em superfluid} regime) to one with
pinning ({\em insulator}) when the phase difference between
adjacent condensates $\Delta \phi(t)$ reaches $\pi/2$. In this
case the relative phases start to run independently with different
velocities, the phase coherence through the array is lost and the
system behaves as an {\em insulator}. The system reaches this
regime for a critical velocity
\begin{equation}
 v_c=\frac{K \lambda}{\hbar}.
\label{cr-vel}
\end{equation}
where $K$ is proportional to the tunnelling rate between adjacent
sites of the optical potential. In a harmonic potential this
corresponds to a critical displacement:
\begin{equation}
\Delta z_{cr}=\frac{\lambda}{2} \sqrt{\frac{2K}{\Omega}}
\label{cr-displ}
\end{equation}
where $\Omega= \frac{1}{2} m \omega_z^2 \big(\frac{\lambda}{2}
\big)^2$ describes the magnetic potential energy of the
condensates. Since the tunnelling rate $K$ depends on $V_0$, the
critical displacement should depend on the height of the interwell
potential: the greater $V_0$ is, the lower the tunnelling rate and
the critical displacement are.

In Fig.~\ref{fig3} we report the DNLS prediction for the critical
displacement (continuous line \cite{thanks}). Effectively, the
theoretical curve well divides the two dynamical regions observed
in the experiment.

\begin{figure}[h]
\begin{center}
\includegraphics[width=10cm]{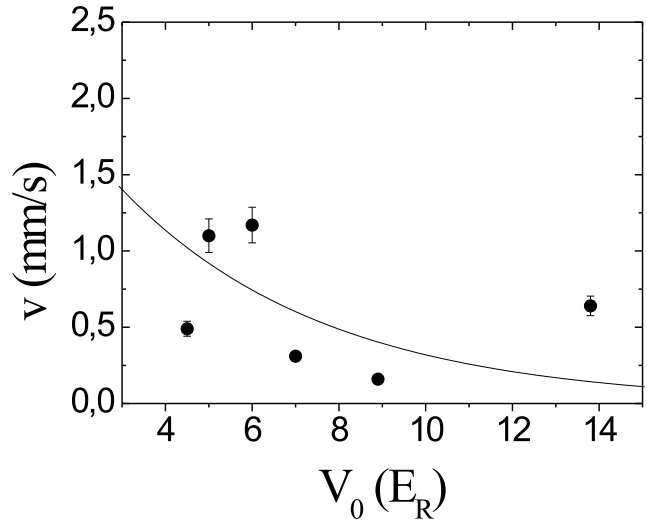}
\caption{Critical velocity as a function of the optical lattice
depth. The experimental results (filled circles) are compared with
with Eq.~(\ref{cr-vel}) (continuous line).}
\label{fig4}
\end{center}
\end{figure}

From the experimental data (open circles in Fig.~\ref{fig3}) we
can also extract the maximum velocity reached by the system before
entering the instability regime. In Fig.~\ref{fig4}, we show the
experimental maximum velocities measured as a function of the
optical potential height, in qualitative agreement with the
theoretical prediction of Eq.~(\ref{cr-vel}) (continuous line).

We believe that the $1D$ model gives a reliable estimate for the
onset of the loss of superfluidity, however it does not
quantitatively describe the degradation of the interferogram in
the present experimental setup. Actually, in the experiment, for
values of displacement just above the predicted $\Delta z_{cr}$ we
could not observe a complete disruption of the interferogram but
rather the appearance of more complex structures within the three
visible peaks. We believe that this could be due to the excitation
of radial modes that are obviously not included in the simple $1D$
model. A possible way to include such modes is to resort to a full
numerical solution of the $3D$ Gross-Pitaevskii equation
\cite{michele}.

In conclusion, we have experimentally observed the transition
between different regimes in an array of Josephson junctions
realized with BECs trapped in a one dimensional optical lattice.
The transition occurs between a \textit{superfluid} and an
\textit{insulator} regime and is accompanied by a loss of
coherence through the array even though each condensate in the
array is still described by a coherent state. The experimental
findings have been compared with the prediction of a one
dimensional theoretical model \cite{smerzi02} based on a discrete
nonlinear Schr\"odinger equation where the transition is due to a
dynamical instability taking place when the eigenfrequencies of
the excitation spectrum become imaginary. This model is found to
qualitatively describe the instability onset which occurs at a
critical value of the displacement or, equivalently, when the
velocity of the wavepacket's center of mass is larger than a
critical velocity dependent on the tunnelling rate through the
optical barriers. We have also observed that close to the
transition the $1D$ model fails to describe the experimental
results where the system behaves like an {\em insulator} but
coherence is still present through the array. We attribute this
behavior to the excitation of radial modes that would require a
full $3D$ simulation of the GPE.

A quantitative comparison between the critical velocity for the
dynamical instability and the speed of sound in the presence of
the optical lattice in the regime of high optical lattice to our
knowledge is still lacking. We think that it will be very helpful
to enlighten the role of this two instability regimes already
discussed for the case of small lattice height in \cite{wu01}.

We are indebted to F.~Minardi and S.~Burger for their help in the
initial stage of the experiment and for several discussions. We
acknowledge A.~Smerzi, A.~Trombettoni, A.~R.~Bishop,
P.~G.~Kevrekidis, M.~Modugno, A.~Tuchman and F.~T.~Arecchi for
many fruitful discussions. Experimental evidences of a similar
instability has also been found in \cite{kasevich}.

This work has been supported by the EU under Contracts No. HPRI-CT
1999-00111 and No. HPRN-CT-2000-00125, by the MURST through the
PRIN 1999 and 2001 Initiative, by the INFM Progetto di Ricerca
Avanzata ``Photon Matter''.


\begin{thebibliography}{<num>}
\bibitem{thanks} The critical displacement has been obtained
from Eq.(\ref{cr-displ}) using for the parameter K the results of
numerical simulation provided by A.~Trombettoni.
\end{thebibliography}
\end{document}